\documentclass[aps,prb,preprint,groupedaddress]{revtex4}
\usepackage{graphicx}
\newcommand{\be}{\begin{equation}}
\newcommand{\ee}{\end{equation}}

\begin{document}
\title{Lattice dilation near a single hydrogen molecule in an 
interstitial channel within a nanotube bundle}
\author{M. Mercedes Calbi, Ari Mizel and Milton W. Cole}
\affiliation{Physics Department, Pennsylvania State University, 
University Park, Pennsylvania 16802}
\date{\today }

\begin{abstract}

We explore the ground state of a single hydrogen molecule within an 
interstitial channel (IC) of a bundle of carbon nanotubes. A previous 
(variational) study found that when many molecules are present, 
comprising a dense fluid, the nanotube lattice is slightly dilated, with a 
1\% relative increase of lattice constant. Although small, that dilation 
doubled the binding energy per molecule inside the ICs. Here, in the 
case of a single particle, the result is an even smaller dilation, localized 
near the particle, and a much smaller increase of the binding energy.

\end{abstract}

\maketitle

\section{Introduction}

The nature of adsorbed phases within a bundle of carbon nanotubes has 
been explored recently with many techniques \cite{hete,oscdiff,osche,bob,
wang,revaldo,will,kan&bill,uptake,boni,rmp,hecond,dil,interc}. The varied motivations 
for this effort include interest in both basic scientific questions 
(e.g., novel phases of matter) and possible applications (gas storage, 
separation and sensing). Most of the theoretical and simulation research 
has assumed that the nanotubes are neither deformed nor displaced by the 
adsorbate. Such assumptions are convenient and might appear to be well 
justified by the small interaction between the gas and tubes (relative 
to the large cohesive energy of the tubes). However, our recent studies 
\cite{dil,interc} found that the behavior of He, Ne and H$_2$ gases within interstitial 
channels (ICs) contradicts that assumption. In particular, the ground state 
of the system corresponds to a slight expansion of the lattice of nanotubes 
in order to accommodate the gas (which is a moderately high density fluid at 
zero line pressure). The resulting expanded lattice provides a much larger 
binding energy for the particles because the interstitial potential of the 
gas is extremely sensitive to the tubes' separation. For example, in the 
case of H$_2$, the binding energy per molecule doubles when a 1\% increase of 
lattice constant occurs. While the exact numbers found in that calculation 
are sensitive to the details of the potential, which is only approximately 
known, a large effect of dilation is expected to be present for all such 
potentials that have been proposed to describe these particles within ICs. 
The reason is that the particles experience strong forces in this environment, 
implying strong reaction forces acting on the neighboring tubes.

One interesting concomitant of this dilation was suggested in the previous 
study. The fact that the binding energy per molecule, $|E_N|/N$, of many 
molecules in 
a dilated lattice is much greater than that of a single molecule in an 
undilated lattice, $|E_1^0|$, suggests that there exists a strongly 
attractive, effective interparticle attraction, mediated by the lattice. 
Here, the superscript $0$ refers to the value in the absence of any dilation. 
If true, this ``mediation hypothesis'' implies that the molecules will 
condense 
to form an anisotropic liquid at and below a high critical temperature $T_c$ , 
an estimate of which 
is provided by the increase in binding energy due to the 
dilation:$(|E_N| -|E_N^0|)/N \approx 300$ K for H$_2$. An analogy occurs in the BCS 
model of superconductivity, in which case the dynamical lattice deformation 
(phonons) mediates an electron-electron attraction, which drives the 
transition. The mediation hypothesis hinges on the implicit assumption that 
the strong binding found for the many-particle system is not present in the 
analogous one-body problem (including dilation). The heuristic idea 
underlying the hypothesis is simple and plausible: a single particle cannot 
deform its environment significantly, so only small dilation occurs in its 
vicinity. If several particles aggregate in the IC, instead, they can share 
the benefit of the lattice deformation without much additional cost in 
lattice energy. Thus, the large increase in binding is a cooperative, 
many-body effect and a high value of $T_c$ is expected.

The present paper is one contribution toward developing and testing this 
hypothesis. We compute the energy of a single H$_2$ molecule in the presence 
of a deformed environment. The procedure employs a Born-Oppenheimer type 
of approximation, fixing the environment with a given deformation and finding the 
hydrogen eigenenergy in that environment. This is also analogous to the 
small-polaron theory employed to treat electrons in alkali-halide crystals 
\cite{kitt}. It may therefore be appropriate to term the entity consisting of the 
hydrogen molecule and its surrounding dilation a ``dilaton''. The net 
energy (tube lattice energy 
+ H$_2$ energy) is minimized with respect to possible deformations.
We evaluate the ground state energy of the dilaton by solving the 
Schr\"odinger equation 
in the presence of an anisotropic lattice distortion, centered on the 
mean position of the impurity. We obtain a result that is consistent with 
the mediation hypothesis: 
the binding energy of a single molecule in the deformed lattice, $|E_1|$, is 
only slightly greater than that in the undeformed lattice, $|E_1^0|$. This 
means that the enhanced binding of the large $N$ system \cite{dil} is 
necessarily a 
consequence of the effective interaction between two or more dilatons. 
Hence, the hypothetical transition should indeed exhibit a high critical 
temperature.

The organization of this paper is the following. In section II, we describe 
the calculation of the elastic energy $E_{latt}$ of the deformed nanotube 
lattice. We compute the potential energy $V$ and the  
H$_2$ energy $E_1$ inside the channel in Section III. Then, we find the 
ground state of the system (lattice + molecule) by minimizing the total 
energy $E=E_{latt} + E_1$ with respect to possible lattice distortions. 
Section IV summarizes our results and comments on its implications for 
the N-body problem.

\section{Nanotube Bulge Energy}

The interstitial channel between three parallel nanotubes will bulge beyond 
its equilibrium size to accommodate a molecule in the channel.  Because of 
the rigidity of the tubes, the bulge will swell gradually along the channel 
axis and then contract again with some long length scale $w$ (Fig.1). The 
deformation will involve long wavelength acoustic modes of the nanotubes; it 
is energetically prohibitive for the unit cells of the tubes to undergo short 
wavelength, optical deformations. It is therefore appropriate to model the 
nanotube unit cells as rigid cylinders of length $c = 2.5$ \AA, so that each 
nanotube consists of a chain of cylinders secured end-to-end by spring forces 
\cite{Mizel}. The forces within each chain of cylinders, as well as the forces 
between chains, 
can be obtained from the Born-von Karman force constants of reference [16] 
provided that nanotubes are assumed to be of the (10,10) variety.

\begin{figure}
\includegraphics[height=2in]{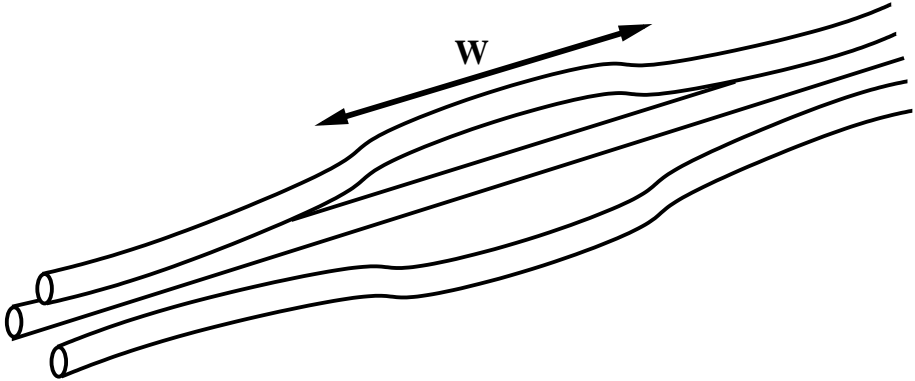}
\caption{Schematic depiction of the nanotube bulge. }
\end{figure}

Our ansatz for the shape of the bulge assumes that the three central 
nanotubes get pushed symmetrically outward a distance $d(z)$ by the molecule 
(Fig. 2).  We adopt a Lorentzian form to describe the magnitude of the displacement 
away from equilibrium

\begin{equation}
\label{lorentzian}
d(z) = \frac{h}{1 + (z/w)^2}.
\end{equation}

Here, $h$ gives the maximum outward displacement of a unit cell while $w$ characterizes the extension of the bulge along the channel axis.

\begin{figure}
\includegraphics[height=3in]{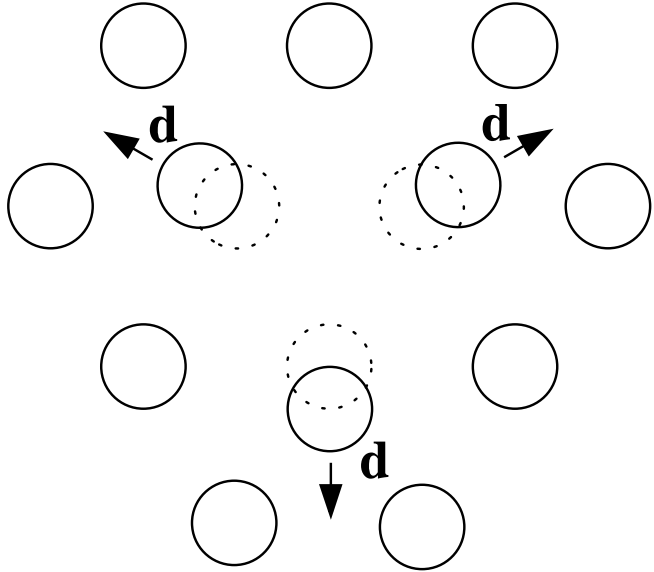}
\caption{Transverse view (at a given $z$) of the deformed IC when the tubes have been pushed a distance $d(z)$ from their equilibrium positions.}
\end{figure}

An internal energy $E_{internal}$ is required for the tubes to bend, which 
is determined by the frequency of transverse acoustic modes $v_{TA}/c$. 
Within the model of Ref. 16 adopted here, it has the value 

\be
E_{internal} = 3 \sum _{l = -\infty}^{\infty} \frac{1}{2} 
M \frac{v_{TA}^2}{c^2} ( d(l + 1) - d(l))^2  
\ee

\noindent Here, $v_{TA} \approx 2 \times 10 ^6$ cm/s is the velocity of the 
transverse acoustic modes that cause a tube to flex like a snake, while 
$M = 8 \times 10^{-22}$ g is the mass of one 
nanotube unit cell, which contains 40 atoms \cite{Mizel}.  An inter-tube energy $E_{3}$ arises as the 
three tubes forming the channel spread apart.  If they are 
members of a nanotube rope, the three tubes will push on $9$ neighoring tubes, 
contributing an additional inter-tube energy $E_{9}$, while the neighboring 
$9$ tubes themselves are assumed to remain in place.  Utilizing the Lorentzian 
form (\ref{lorentzian}) in the energy function of Ref. 16, 
one computes the deformation energy of the tubes as a function of the channel 
bulge shape:

\be
E_{latt}(h,w) = E_{internal} + E_3 + E_9
\ee

\section{Hydrogen dilaton energy}

As in most studies of hydrogen near nanotubes, we ignore the internal 
degrees of freedom of the molecule and focus on the motion of its center of 
mass. We compute the ground state energy $E_1(h,w)$ of the molecule 
in the deformed channel $E_1(h,w)$ by solving the Schr\"odinger equation

\be
\left[-\frac{\hbar^2}{2m}\nabla^2 + V(r,z)\right]\psi(r,z)=E_1 \psi(r,z)
\ee

using the diffusion method. Here, $V(r,z)$ is the potential energy due to 
the deformed tubes. Since the deformation of the tubes is expected to vary 
over a long distance, we evaluate the potential using a local approximation 
(confirmed a posteriori) in which the potential is that of a uniform system 
having the local value of the function $d(z)$: $V(r,z)\equiv U_{d(z)}(r)$, 
where $r$ is the radial distance from the center of the
 channel. The potential energy $U_{d(z)}(r)$ is obtained by summing the 
contribution from
 the three tubes surrounding the channel and according to the nanotube 
potential model 
described in Ref.[9]. In this way, we obtain the total energy as:

\be
E (h,w)=E_{latt}(h,w) + E_1 (h,w)
\ee 

The ground state of the system is obtained by minimizing this expression 
with respect to the deformation parameters $h$ and $w$. Figure 3 shows the 
function $E (h,w)$ and the presence of a minimum $E_{min} = -282.5$ K at 
$h=0.005$ \AA~ and $w=8$ \AA. This represents a binding energy increase 
of 1 K with respect to the case of non-deformed tubes. The 
 deformation is extremely small in the transverse direction ($h/a \approx$ 
0.0003, where $a=17$ \AA~ is the unperturbed lattice constant) and it extends 
over a much larger distance ($\approx 16$ \AA) along the axis of the channel. 

The probability density $|\psi|^2$ and the external potential $V$ for the minimum 
energy configuration are plotted in Figure 4. The molecule's localization in 
the transverse direction is very similar to that found in the undilated case, 
with a root mean square transverse displacement $\sqrt{\langle r^2 \rangle}
\approx 0.2$ \AA.The localization in the 
direction parallel to the axis has a much larger spread of the wave function, 
$\sqrt{\langle z^2 \rangle}
\approx 1.6$ \AA. Note that the longitudinal spread of the wave function is 
much smaller 
than the spread of the deformation that is responsible for the localization. 
This is evident in Figure 4: the equipotential lines in Fig. 4(a) 
show more anisotropy than the constant probability lines of Fig. 4(b). 
 The line shapes are consistent with a simple model calculation in which 
one assumes an anisotropic harmonic oscillator potential with radial and 
axial frequencies in the ratio $\omega _r / \omega _z = 150.$

\begin{figure}
\includegraphics[height=4in]{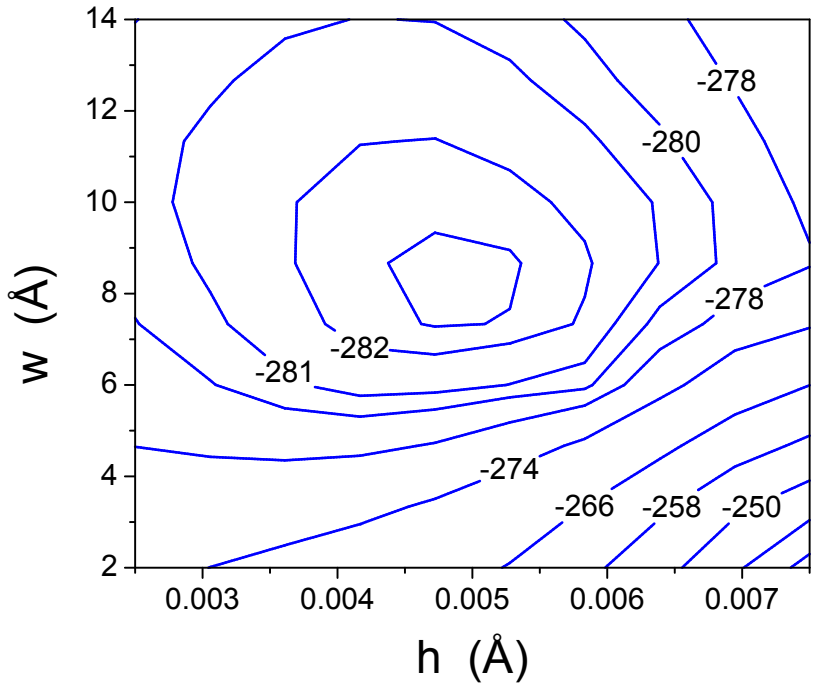}
\caption{Ground state energy $E_{latt}+E_1$ (in K) as function of the deformation parameters $h$ and $w$. A minimum $E=-282.5$ K occurs for $w=8$ \AA~ and $h=0.005$ \AA.}
\end{figure}

\begin{figure}
\includegraphics[height=5in]{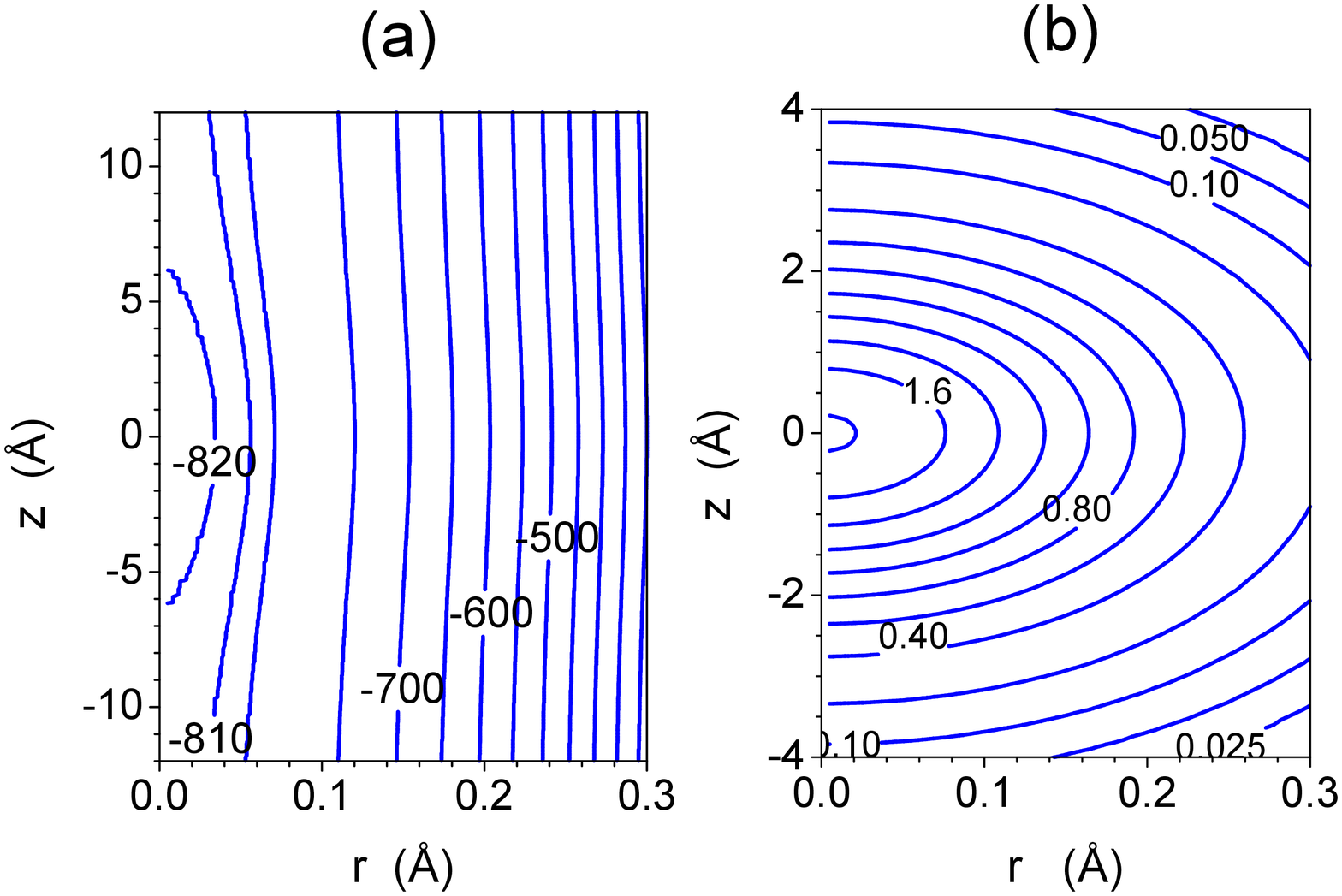}
\caption{Dilation properties for the lowest energy case. (a) External potential energy (in K) due to the deformed tubes as function of the radial distance $r$ and the distance $z$ along the axis of the channel. From left to right, the first three isopotential contours correspond to -820 K, -810 K and -800 K, followed by contours of increasing energy every 50 K. (b) Probability density (normalized in units of \AA$^{-3}$) of a hydrogen molecule for the same case as (a). From 1.8 \AA$^{-3}$ to 0.2 \AA$^{-3}$, the contours are 0.2 \AA$^{-3}$ apart. Note the differing scales on the ordinates of the two figures.}
\end{figure}

\section{Summary}

We have investigated the ground state properties of the hydrogen dilaton,
consisting of a localized molecular wave function plus the accompanying lattice
deformation. As assumed from the start, the rigidity of the tubes implies that
the longitudinal extent of the deformation is large compared to the other
length scales in the problem. The ground state energy obtained by minimizing
the total energy is just 1 K below the ground state energy of  a molecule
interacting with the rigid nanotube lattice; this value is the net result of a
10 K increase in lattice energy and an 11 K decrease in energy of the molecule 
due to the deformation.
While the exact values of these energies are sensitive to the assumptions in
the potential energy calculation, their small magnitude (compared to the
binding energy  of the many-body system) should be a robust conclusion of our
study. This leaves open the question posed in the introduction: how does one
account for the huge cohesive energy found in our previous study of the
many-body system?

That  large cohesive energy (about 200 K) is necessarily a consequence of
inter-dilaton couplings: 2 body, 3 body,... One may speculate, for example, 
about whether a
dilaton dimer, consisting of two molecules plus their deformation, has a much
lower energy than two isolated dilatons. This might well be the case because
a second molecule can exploit the rather extended deformation we have found for
a single molecule. This argument would imply that the binding energy of the
dimer is at least 10 K because the dimer need not produce a deformation that is
longer than that shown in Fig. 4 for one molecule.

We intend in future work to explore this dimer problem by a technique similar to
that used here, which is essentially a variational, strong-coupling method. If
we find greatly enhanced cohesion in that problem, it will imply that the
two-dilaton interaction is strongly attractive, with a resulting high
temperature critical transition.

\begin{acknowledgements}

We are grateful to Gerry Mahan for many helpful discussions. A. M. acknowledges 
the support of a 2001 Packard Fellowship for Science and Engineering. M.M.C and 
M.W.C. acknowledge support from the National Science Foundation.

\end{acknowledgements}

\end{document}